# Effect of Negative Chemical Pressure on the Prototypical Itinerant Magnet MnSi

C. Dhital,[1,*] M.A. Khan,[1] M. Saghayezhian,[1] W. A. Phelan,[1] D. P. Young,[1] R. Y. Jin,[1] and J. F. DiTusa[1,**]

[1]Department of Physics and Astronomy, Louisiana State University, Baton Rouge, LA 70803

**Abstract:**

The evolution of the magnetic and charge transport properties of the itinerant magnetic metal MnSi with the substitution of Al and Ga on the Si site is investigated. We observe an increase in unit cell volume indicating that both Al and Ga substitutions create negative chemical pressure. There are substantial increases in the Curie temperature and the ordered moment demonstrating that the substitutions give the magnetism a stronger itinerant character. The substitutions also increase the range of temperature and field where the skyrmion phase is stable due to a change in the character of the magnetism. In contrast to the behavior of pure MnSi and expectations for the intrinsic anomalous Hall effect, we find a significant temperature dependence to the magnitude and sign of anomalous Hall conductivity constant in Al or Ga substituted samples. This temperature dependence likely reflects changes in the spin-orbit coupling strength with temperature, which may have significant consequences on the helical and skyrmion states. Overall, we observe a continuous evolution of magnetic and charge transport properties through positive to negative pressure.

## 1. Introduction:

Since the discovery of magnetic skyrmion lattices in *B20* structured materials in 2009 [1], suggestions of applicability of this phenomenon to magnetic storage devices have been made due to the mesoscopic size of the magnetic textures and the small currents necessary to influence them [1,2]. In order to realize this vision, much greater control over the materials and spin textures must be achieved. Features such as the magnetic transition temperature, skyrmion lattice constant, helicity, and stability, as well as the coupling to the charge and lattice degrees of freedom, require exploration via changes to the chemistry and application of physical fields and pressure. Here, we investigate the effect of chemical substitutions of Al and Ga on the Si site of the most well-studied of the *B20* materials, MnSi. These substitutions have the effect of increasing the lattice constant by effectively applying a negative chemical pressure, changing the carrier density, and increasing the disorder. We explore the variations to the magnetic and charge transport properties that these substitutions cause to this prototypical itinerant helimagnet.

MnSi is perhaps the most thoroughly studied of the class of itinerant magnetic materials, a history that includes detailed investigations of the effect of hydrostatic (positive) pressure [3-11]. Neither the Stoner theory of itinerant moments nor the Heisenberg theory of localized moments can properly describe the magnetic behavior [12, 13] of this model system. At ambient pressure MnSi undergoes a transition from paramagnetic to helimagnetic at a temperature $T_C$=29 K with a long helimagnetic wavelength (180 Å) propagating along the [111] crystallographic axis [13, 14]. The magnetic susceptibility, $\chi$, follows a Curie-Weiss behavior at high temperature giving an effective moment of 2.2 $\mu_B$/Mn, whereas the saturation moment at low temperature is only 0.4 $\mu_B$/Mn, placing MnSi clearly in the category of weak itinerant magnets [13, 14]. This behavior has been modeled by taking into account an intermediate correlation strength to go along with the comparable bandwidth and exchange correlation energies [15-16]. In MnSi both the saturation magnetic moment and $T_C$ decrease progressively with external positive pressure which is accompanied by a transition from Fermi-liquid to non-Fermi liquid (NFL) charge transport [3-5]. The NFL behavior extends over a wide range of pressure and coexists with a partial magnetic order giving an excellent example of NFL behavior in the absence of quantum criticality [3-11]. In addition, a quantum phase transition can be accessed through chemical substitution. Fe and Co substitution on the Mn site ($Mn_{1-p}Fe_pSi$ and $Mn_{1-r}Co_rSi$) drives the system progressively toward a quantum critical point where the critical concentration depends directly upon the electron count; that is the critical substitution level, $x_c$, for Co is only half of that found for Fe substitution [17-19].

Over the past few years, there has been enormous interest in MnSi associated with the discovery of a nontrivial topological magnetic phase called the *A*-phase or Skyrmion lattice phase over a small range of magnetic fields just below $T_C$ [1]. The antisymmetric Dzyaloshinskii-Moria (DM) interaction (*D*) arising from broken inversion symmetry of this chiral structure favors a perpendicular spin alignment that competes with the Heisenberg exchange interaction (*J*). Although *D* is small compared to *J*, it is sufficient to cant the magnetic moments to produce a helical magnetic structure with a long wavelength ($\lambda \sim J/D$) [1]. Just below $T_C$ thermal fluctuations stabilize the skyrmion lattice phase at small fields [1, 2, 20-23]. In addition to these fundamentally important aspects, there is also interest in its potential use in spintronic devices. The interest stems from the realization that skyrmions are a collective excitation that respond as a single particle, thereby requiring a much smaller current to control as compared to conventional magnetic memory devices [2, 21].

Several theoretical and experimental studies have been performed in an attempt to understand the electronic and magnetic behavior of MnSi in the presence of external hydrostatic pressure and chemical substitution. However, most of these focused either on creating positive pressure [3-11] or on chemical substitutions on the Mn site [17-19]. Given its interesting properties, it is natural to ask how the electronic and magnetic behaviors vary in the presence of negative pressure. Here, we employ chemical substitution of elements having a larger ionic radius as an effective way to explore the effect of negative chemical pressure. There have been relatively few studies focusing on chemical substitution on the Si site [24, 25]. One of these reported a significant increase in $T_C$, by 10 K, with less than 1% Ge substitution for Si [24]. The extreme sensitivity of $T_C$ to chemical substitution was thought to be due to an increased localization of the magnetic moments stemming from an increased negative chemical pressure. However, in a separate study [25], no significant change in $T_C$ was observed even with 10% Ge substitution. Here, we present a systematic investigation of the evolution of the electrical and magnetic behavior of MnSi with Al and Ga substitution, in part, to clarify these contradictory results. The choice of these particular substitution elements allows the investigation of the role of negative chemical pressure and the effect of non-isoelectronic substitution on the crystal structure, ac and dc magnetization, resistivity, and Hall effect. Despite the indirect role of Si on the magnetism of this compound, we find that these substitutions significantly increase the transition temperature and the saturation moment in addition to stabilizing the skyrmion lattice phase over a wider temperature and field range while preserving all the qualitative features of magnetism found in the complex MnSi magnetic phase diagram [1]. Our results indicate that the negative chemical pressure is the primary factor for such a change in magnetic behavior consistent with previous results [24]. We support our findings by performing positive pressure experiments, which reverse the effect produced by negative pressure. It appears as if pressure, both positive and negative, acts as an effective knob for tuning the magnetic properties of MnSi. In addition, we find significant variation in the anomalous Hall effect (AHE) which does not conform to the usual understanding of the intrinsic AHE.

## 2. Experimental Details

Polycrystalline samples were obtained by arc melting high purity elements in an inert argon atmosphere. These samples were annealed at 1000 ºC for approximately 3 days in sealed quartz tubes. Single crystals of MnSi$_{1-x}$Al$_x$ ($x$<0.04) and MnSi$_{1-y}$Ga$_y$ ($y$<0.01) were obtained by loading these polycrystalline pellets inside graphite tubes and employing the modified Bridgman method in a RF furnace under a flowing argon environment. Attempts to synthesize phase pure single crystals for higher Al or Ga concentrations at ambient pressure were unsuccessful indicating solubility limits for these substitutions. Additional MnSi$_{1-y}$Ga$_y$ samples were grown using Ga as a flux as described in Ref [26]. Mn$_{1-p}$Fe$_p$Si (p<0.1) and MnSi$_{1-z}$Ge$_z$ (z < 0.1) polycrystalline samples were prepared by arc melting high purity elements in an inert argon atmosphere. The phase purity and crystallinity of our samples were checked by employing both powder and single crystal x-ray diffraction.

Having established the crystal structure, the absence of second phase formation, and the systematic variation of lattice constants in our samples, we next employed electron beam techniques to determine the stoichiometry accurately. For chemical analysis we have used JEOL JSX-8230 Electron Microprobe located at Shared Instrumentation Facility (SIF) at Louisiana State University. This instrument allows simultaneous measurement via Wavelength Dispersive Spectroscopy (WDS) and Energy Dispersive Spectroscopy (EDS) techniques. The specific implementation of the microprobe allows microscopic regions 0.5-1µm of the sample to be probed. The procedure followed included selecting at least two crystals (or polycrystalline samples in the case of arc-melted samples) from each synthesis batch and polishing them so that they are optically smooth. Each of these samples was scanned at 5 or more different areas within a region of about 500 µm. We report the average values of the concentrations determined which had a spread of less than Δ$x$=0.002 in absolute value (see Fig. 1d). In addition, some sample surfaces were coated with carbon to check for problems due to possible low surface conductivity. No such problems were found.

Magnetization measurements, both ac and dc, were carried out in a Quantum Design 7-T MPMS SQUID magnetometer. Unless otherwise stated, ac susceptibility measurements were performed at a frequency of 100 Hz with an ac driving amplitude of 1 Oe. High pressure magnetization measurements were performed by using a Cu-

Be pressure cell with Daphne 7373 oil as the pressure transmitting medium. The pressure in the cell was determined from the change in the superconducting transition temperature of Pb wire pressurized along with the sample. Resistivity and Hall effect measurements were performed using standard lock-in techniques at 19 Hz with currents of 1 mA or less with any self-heating carefully monitored. All samples had a rectangular cross-section and were thinned by polishing down to ~150 µm. Thin Pt wires were attached to four Epotek silver epoxy contacts with an average spacing between the voltage probes of 0.4 mm. Hall resistivity data were corrected for any misalignment of leads by symmetrizing the data collected at positive and negative fields.

## 3. Experimental Results

The results of dc magnetic susceptibility and magnetization, *M*, measurements performed on single crystal samples are presented in Fig 1. The sharp change in $\chi$ as function of temperature defines the transition temperatures as defined in Fig S1 of the Supplementary Materials (SM)[27]. From Fig 1a it is clear that $T_C$ increases upon substitution of either Ga or Al. Fig 1b presents *M* as function of field, *H*, at 5 K for these same samples. The saturated magnetic moment also observed to systematically increase with substitution. Fig 1c represents a Curie-Weiss fit to the high temperature part of the inverse susceptibility, and Fig 1d represents the variation of effective moments, $m_{eff}$, Weiss temperature ($\theta$), and $T_C$. It is clear that all of these samples exhibit Curie-Weiss behavior [Fig 1c] with increased $\theta$ and effective moment, $m_{eff}$, that track the increases in $T_C$ [Fig 1d].

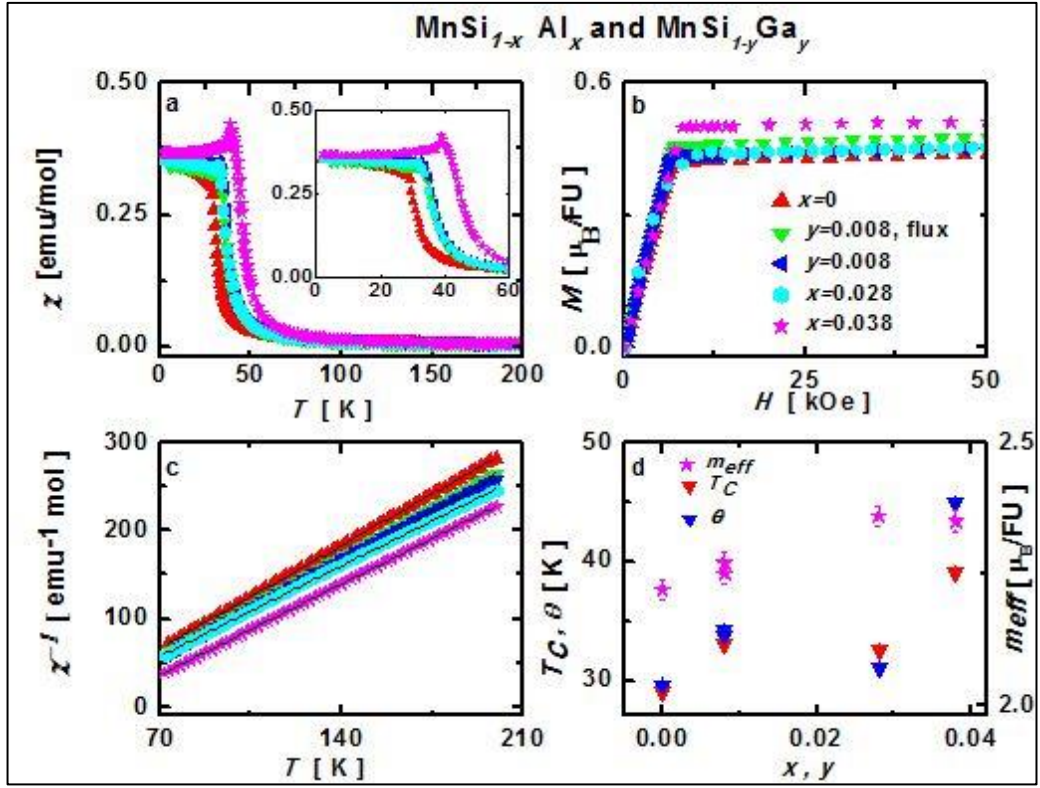

**Fig 1** : dc magnetization of MnSi$_{1-x}$Al$_x$ and MnSi$_{1-y}$Ga$_y$ single crystals. (a) dc susceptibility, $\chi$, as a function of temperature, $T$, for a magnetic field, $H$=1 kOe. Symbols are the same as in frame b. Inset: Low temperature $\chi$. (b) dc magnetization, $M$, as function of $H$ at 5 K (c) $\chi^{-1}$, vs. $T$ displaying Curie-Weiss behavior above the transition temperature, $T_C$. Symbols are the same as in frame b. Lines are fits of the Curie-Weiss form to the data. (d) Variation of $T_C$, Weiss temperature, $\theta$, (left axis) and Curie moment, $m_{eff}$, (right axis) as function of $x$ or $y$.

The correlation between the lattice parameter of our crystals and $T_C$ is presented in Figure 2. The lattice parameter increases in response to the substitution resulting in a negative chemical pressure within the crystal. Thus, Figs. 1 and 2 demonstrate that an increase in $T_C$, the saturation magnetization, Curie moment, and the Weiss temperature coincide with increases in the lattice parameter for both Al and Ga substitutions. There are two probable causes for the magnetization changes we observe as the substitutions not only produce an increased lattice constant but also reduce the valance electron density. To probe the relative importance of these changes we have synthesized polycrystalline samples of MnSi$_{1-z}$Ge$_z$ [$z$<0.01] and Mn$_{1-p}$Fe$_p$Si [$p < 0.1$] as well as a number of Al substituted polycrystalline samples for comparison to our crystalline materials. Unlike Al or Ga substitution, Ge substitution is isoelectronic. Fig 2a demonstrates the variation of $T_C$ as function of lattice parameter for Ga, Al, Ge, and Fe substituted samples where a unified dependence of $T_C$ on lattice parameter is apparent for Ga, Al, and Ge substitution.

For the case of Fe, $T_C$ decreases faster than the trend followed by the other substitutions, perhaps reflecting the effect of a significantly reduced magnetic moment in addition to positive pressure [28]. The similarity found for Ga, Al, and Ge substitutions indicates that the change in lattice parameter is the primary reason for the change in $T_C$ that we observe.

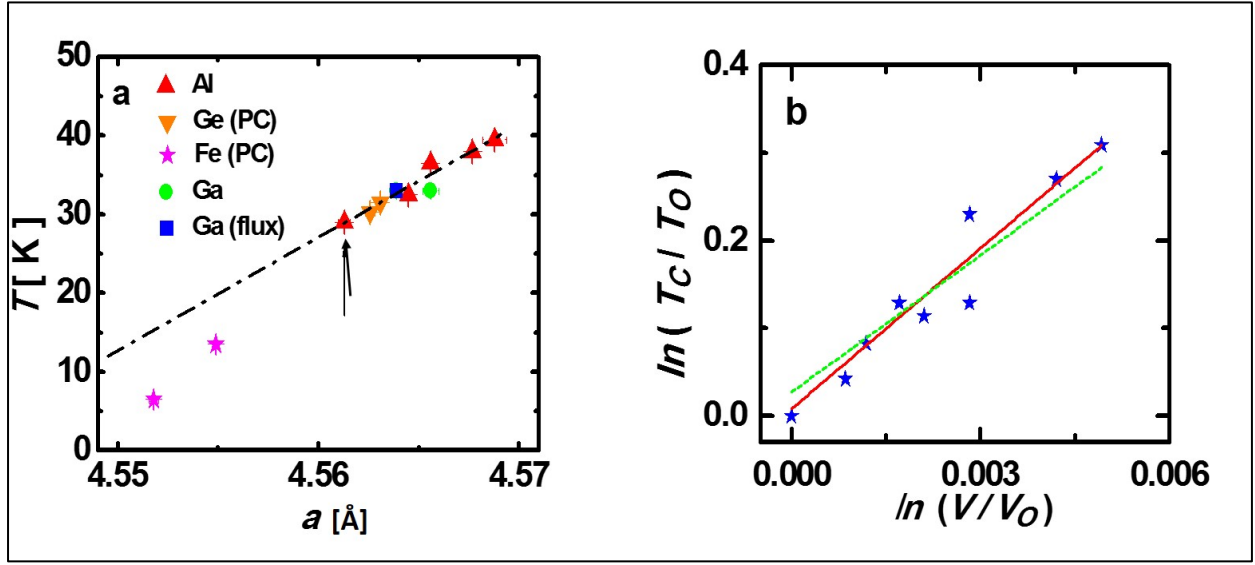

**Fig 2** Curie temperature variation with lattice expansion (a) Curie temperature, $T_C$, as function of lattice parameter, $a$, for MnSi, $MnSi_{1-x}Al_x$, $MnSi_{1-y}Ga_y$, $MnSi_{1-z}Ge_z$ and $Mn_{1-p}Fe_pSi$ crystals. The dashed line is a guide to eye demonstrating linear behavior. Black arrow indicates the lattice parameter of nominally pure MnSi (b) Variation of $ln(T_C/T_O)$ with the logarithm of the volume normalized to $V_O$, $ln(V/V_O)$, corresponding to the data in (a) (data for samples with Fe substitution omitted). $T_o$ and $V_o$ represent the Curie temperature and volume of the unit cell for pure MnSi respectively. The red solid line is a linear fit to the data with a slope of 62+/- 7 determined by the fitting procedure, while the green dashed line is a linear fit using a constant slope of 50 +/-2 as given in Ref [8].

Several previous studies that focus on the change in magnetic properties due to positive hydrostatic pressure have explained the decrease in the magnetic moment and the transition temperature in terms of the increased delocalization of the electron wave functions [3-11]. Our observation of increased $T_C$ with negative chemical pressure in the current investigation is consistent with the dependence found in the previous studies with positive pressure. This is demonstrated in Fig 2b which displays the change in $T_C$ with changes to the volume of the unit cell, which we parameterize via a linear relationship. The slope of the best-fit line to these data is 62±7, which is close to, but slightly outside of the error of the previously reported 50±2 obtained using positive hydrostatic pressure [8] suggesting that our results are primarily an extension of the pressure experiments toward negative chemical pressure.

We have also performed high-pressure magnetization measurements on two representative crystalline Al and Ga substituted samples, Fig 3, to further confirm pressure effect on the magnetic properties of this system. Consistent with previous high pressure studies, the transition temperature decreases with pressure along with a slight reduction in the saturated magnetic moment. The pressure dependence of the Ga flux grown sample is more dramatic at a somewhat larger applied pressure ($P$=1.4 GPa), a value close to the critical pressure needed to completely suppress long range magnetic order in pure MnSi [3,8,9]. In order to compare the effect of pressure on our chemically substituted samples to that of pure MnSi reported in the literature, we calculate the expected change in Curie temperature for our MnSi$_{0.962}$Al$_{0.038}$ sample assuming it has the same bulk modulus and pressure dependence as pure MnSi. For an external pressure of $P$=1.25 GPa, we expect a decrease in the Curie temperature of $\delta T_C$=13.5 K for the x=0.038 sample, a somewhat larger change than the $\delta T_C$=11 K that we observe. In addition, the relative change in the saturation moment observed in Fig. 3 is not as pronounced as that in the presence of hydrostatic pressure in the case of MnSi. Although it is clear that negative pressure is the dominant reason for the changes to the magnetic properties with the chemical substitutions investigated here, there are some small discrepancies when compared to the case of hydrostatic pressure applied to MnSi. Such discrepancies may stem from the combined effect of a change in carrier density and the increased crystalline disorder due to the chemical substitution.

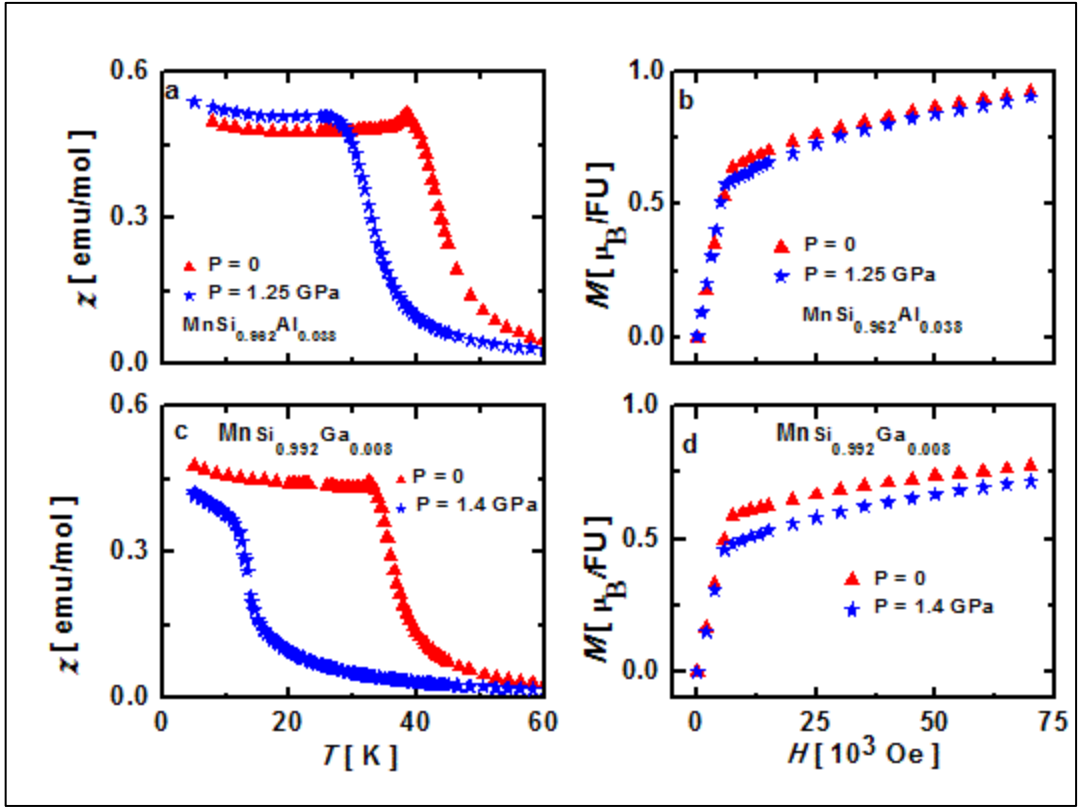

**Fig 3** Pressure Dependent Magnetism. (a) Magnetic susceptibility, $\chi$, as function of temperature, $T$, at pressures, $P$, of 0 and 1.25 GPa for $MnSi_{0.962}Al_{0.038}$ (b) Magnetization, $M$, vs. magnetic field, $H$, at 5 K with $P=0$ and 1.25 GPa for $MnSi_{0.962}Al_{0.038}$ (c) $\chi$ as function of $T$ at 0 and 1.4 GPa for $MnSi_{0.992}Ga_{0.008}$ (synthesized employing the flux method) and (d) $M$ vs. $H$ at 5 K and $P=0$ and 1.4 GPa for $MnSi_{0.992}Ga_{0.008}$.

As mentioned in the introduction, one important feature of this system is that it hosts a topologically non-trivial magnetic structure called the skyrmion lattice phase over a small temperature range below $T_C$ for a limited range of magnetic fields. We performed ac susceptibility measurements to identify the magnetic fields where the skyrmion lattice is stable in both pure, as well as Al and Ga substituted samples. Fig 4 shows the ac susceptibility for one such sample across its transition temperature. We observe four distinct regions in the real part of the ac susceptibility similar in magnitude and appearing at similar fields as in nominally pure MnSi. There are four different characteristic fields (labeled $H_1$ through $H_4$) indicated in Fig. 4. For $0 \leq H \leq H_1$, the susceptibility increases almost linearly with field indicating a progressive change from helical to a conical magnetic phase. For $H_1 \leq H \leq H_2$ investigations of MnSi and $Mn_{1-p}Fe_pSi$ indicate a complete conical phase. For $H_2 \leq H \leq H_3$ there is a dip in the real part of the susceptibility between two distinct peaks in the imaginary part. This represents the slow movement of a massive magnetic object and corresponds to the *A*- phase or skyrmion lattice phase. For $H_3 \leq H \leq H_4$, $\chi_{ac}$ returns to the same value as below $H_2$ so that we assume the system returns to the conical phase in agreement with the behavior of nominally pure MnSi. Finally, for $H > H_4$ the system enters into a field polarized, ferromagnetic-like phase. The exact values of these fields depend somewhat on the sample history as well as sample shape (demagnetization factor) and relative orientation with respect to magnetic field.

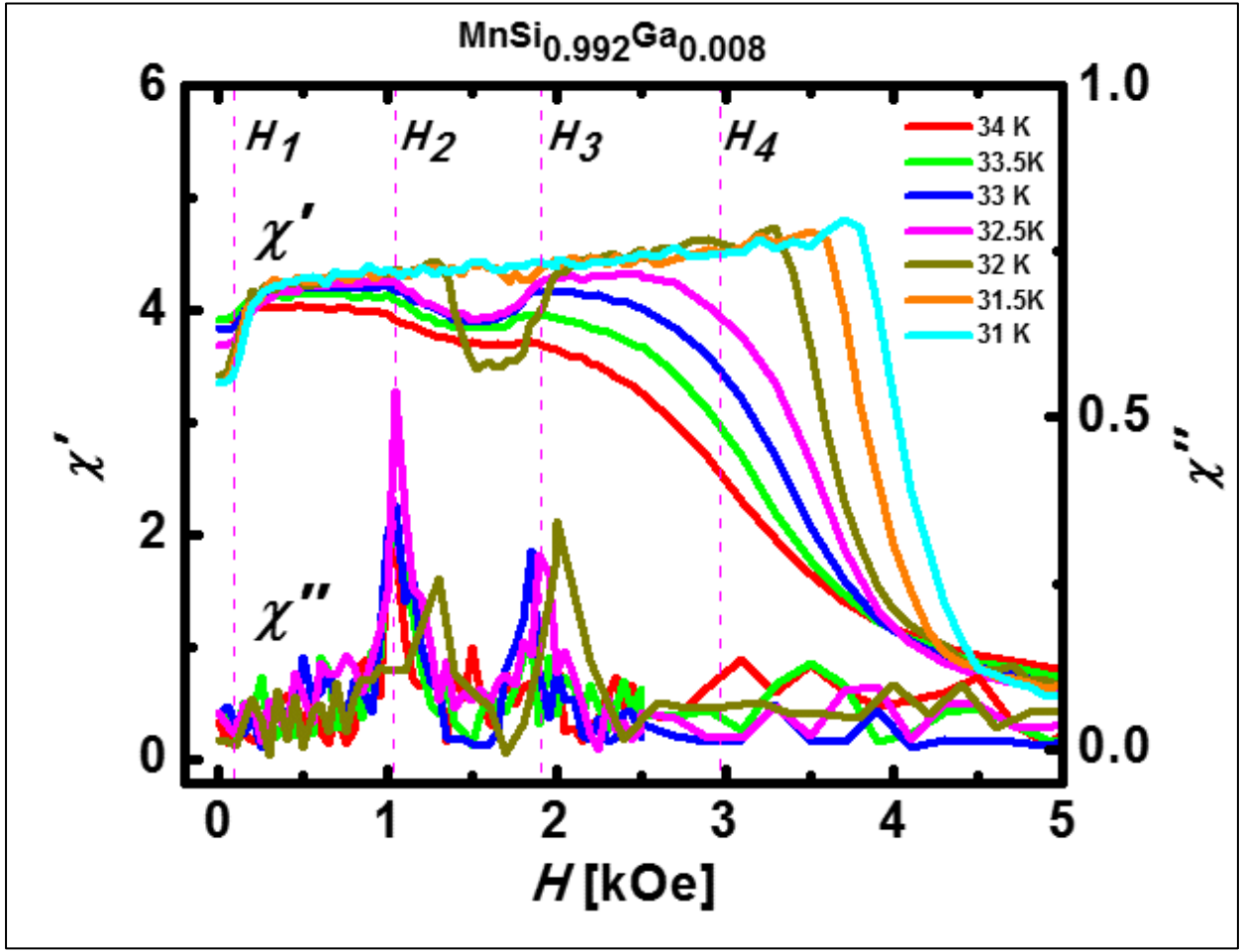

**Fig 4** Real part ($\chi$', left axis) and imaginary part ($\chi$", right axis) of the ac susceptibility as function of dc magnetic field, *H*, at different temperatures for one representative sample, MnSi$_{0.992}$Ga$_{0.008}$ (Bridgman method). As an illustration, the four different fields $H_1$, $H_2$, $H_3$, $H_4$ corresponding to ac scan at 32.5 K (magenta curve) are indicated as described in text.

By performing ac susceptibility measurements for pure, Ga substituted, and Al substituted samples at several temperatures just below their corresponding transition temperatures (Fig S3, Supplementary materials), we map the magnetic phases in the region of the *A*-phase for these samples as presented in Figure 5. Although the absolute value of the transition temperature increases with substitution, the phase diagram remains qualitatively similar. The value of the critical field depends upon the periodicity of the magnetic structure, which in turn is governed by the ratio of $D$ and $J$ [29]. Neutron scattering measurements are necessary to correlate such parameters. However, from these data, we can conclude that the overall magnetic phases, including the skyrmion lattice phase, are stable against Al and Ga chemical substitution giving a qualitatively similar magnetic phase diagram. We also observe that the temperature range over which the skyrmion phase is stable increases with doping concentrations $x$ or $y$.

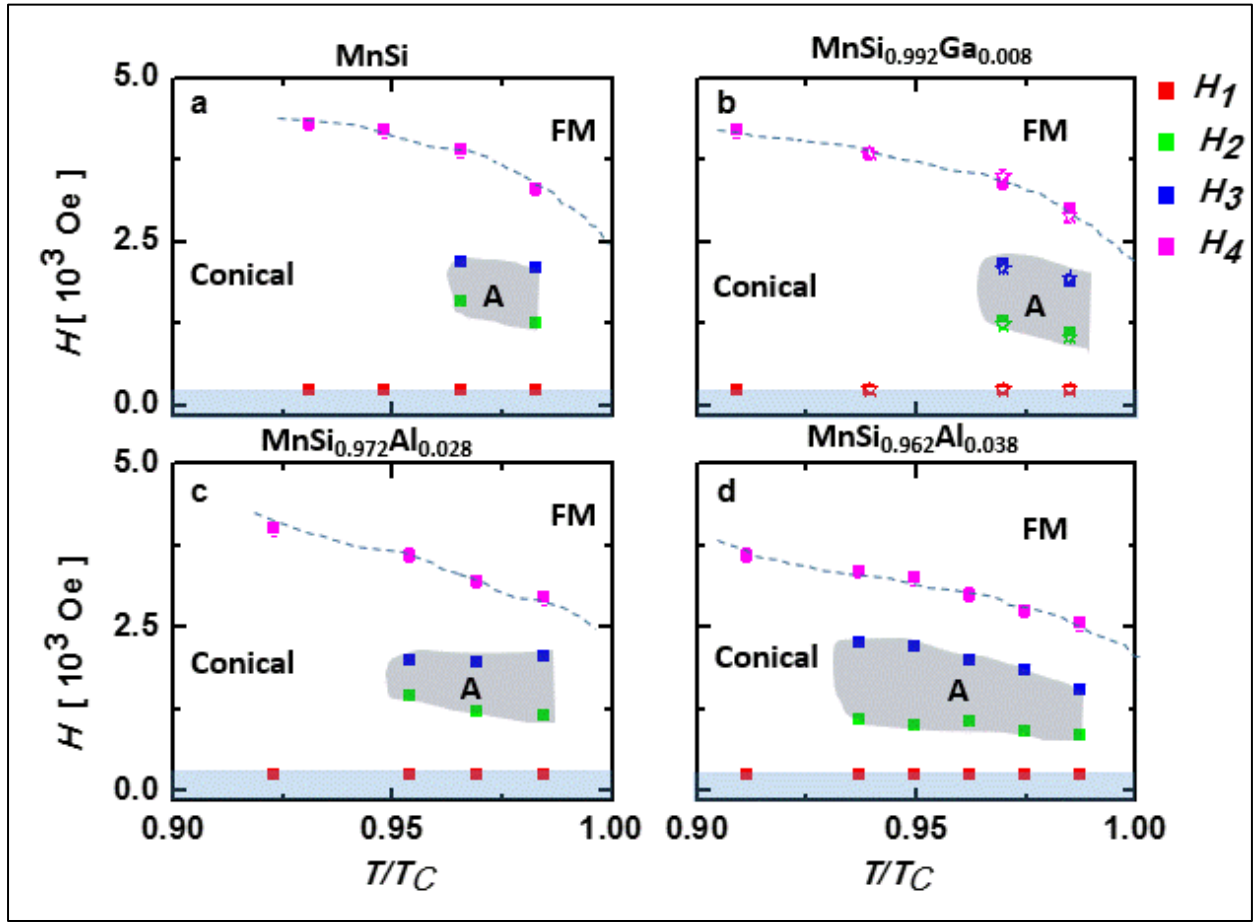

**Fig 5** Magnetic phase diagram for (a) MnSi (b) $MnSi_{0.992}Ga_{0.008}$ (c) $MnSi_{0.972}Al_{0.028}$ (d) $MnSi_{0.962}Al_{0.038}$. In each diagram, the vertical axis represents 4 critical fields $H_1$, $H_2$, $H_3$ and $H_4$ as defined in the text and in Fig 4. The horizontal axis represents temperature, $T$, divided by the corresponding magnetic transition temperature, $T_c$. In frame (b), rectangles represent critical fields for the Ga substituted crystal grown using the Bridgman method and stars represent that for a flux grown sample. The crystals were mounted in a random crystallographic direction relative to the magnetic field.

The resistivity, $\rho_{xx}$, of four samples is presented in Fig. 6. All samples show metallic behavior with only moderately small increases due to the doping induced disorder. Each displays a sudden downturn in resistivity with cooling close to their corresponding $T_C$'s as expected for itinerant magnetic materials where fluctuation scattering is significant near $T_C$. $d\rho_{xx}/dT$ for $x$=0.038 is slightly smaller than rest of the samples which can arise either by disorder, as reflected in the resistivity ratio, RRR, or by a change in the spin fluctuations above $T_C$, which is consistent with a higher $T_C$ and larger saturated magnetic moment, or by a combination of these.

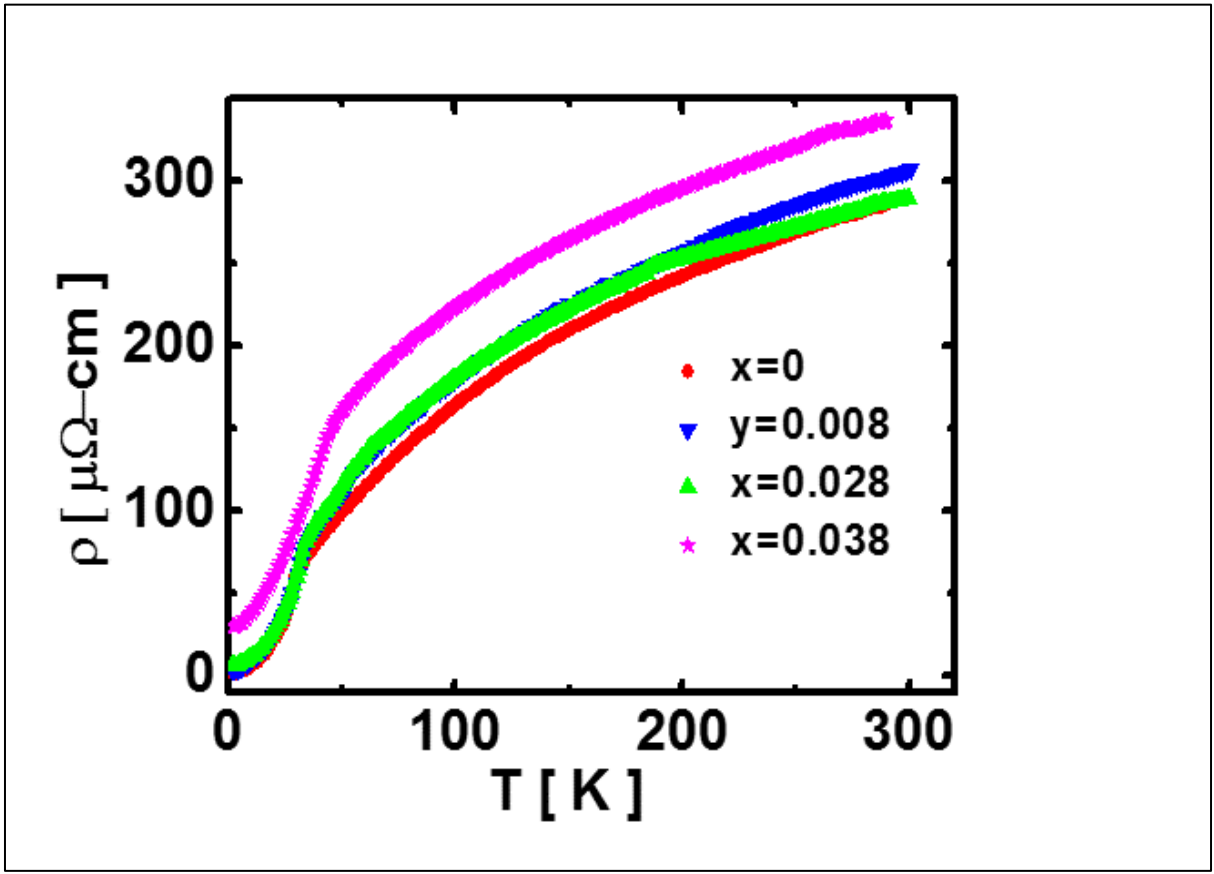

**Fig 6** Charge Transport: Resistivity, $\rho_{xx}$ vs. temperature, $T$.

For a magnetic metal, the analysis of Hall resistivity, $\rho_{xy}$, requires consideration of both normal and anomalous contributions that are parameterized by the relation $\rho_{xy} = R_0 H + 4\pi R_S M$ [30]. The first term, $R_0 H$, with $R_0$ being the ordinary Hall constant represents the ordinary Hall effect that varies linearly with magnetic field and inversely with carrier density in the simplest interpretation. The second term, $4\pi R_S M$, where $R_S$ is the anomalous Hall constant, represents the anomalous contribution. $R_S$ depends on the longitudinal resistivity, $\rho_{xx}$, with the functional form dependent on the dominant anomalous Hall mechanism, previously determined to be intrinsic for MnSi below $T_C$ [18,22,30-34]. The intrinsic anomalous Hall effect, where $R_S \sim \rho_{xx}^2$ is thought to be due to Berry phase curvature in the electronic structure, whereas the high-$T$ anomalous behavior has been reported to change, such that $R_S \sim \rho_{xx}$ [33]. In addition, just below $T_C$, where the system is in the skyrmion lattice or $A$-phase, a third term designated the topological Hall effect has been demonstrated to be present[35]. This topological Hall effect has also been shown to be enhanced due to impurities or pressure [35]. In this manuscript we do not focus on the temperature region just below $T_C$ and, hence, do not consider the topological contribution in our Hall effect analysis. Fig. 7 presents the variation of $\rho_{xy}$ as a function of $H$ and $T$ for four different samples. From Fig 7 (a), (b), and (c), it is clear that $x=0$, $y=0.008$ and $x=0.028$ samples have a qualitatively similar magnetic field and temperature dependent behavior to MnSi. However, our $x=0.038$ sample displays a low temperature $\rho_{xy}$ that indicates a change in sign of the AHE as demonstrated more clearly in inset of Fig 7d.

We have performed the standard analysis of our Hall data as shown in Figure 8 to determine the normal and anomalous Hall coefficients that are presented in Fig. 9. The linear dependence of $\rho_{xy}/H$ as a function of $M/H$ in Fig. 8 has allowed the accurate determination of both $R_0$ and $R_S$ from linear fits to the data. The change in sign of the AHE for our x=0.038 sample is also demonstrated in this figure for two crystalline samples (s1 and s2) grown at different times. From Fig. 9 a and b, it is clear that there is a systematic increase in $R_O$ with dopant density at both 5 K and 25 K. X-ray Photoemission Spectroscopy measurements were unable to detect any change in Mn oxidation state due to the small changes to the charge density expected at this level of chemical substitution (see SM Fig. S2) [27]. Similar to the case of MnSi, Al and Ga substituted crystals display an increased $R_0$ with temperature up to $T_C$ (Fig. 9b). Fig 9c presents the variation of $R_S$ as function of temperature for four different samples. For $x=0$, $y=0.008$ and $x=0.028$, $R_S$ is near zero at 5 K consistent with the small low temperature resistivity of these samples. At higher temperatures, $R_S$ becomes more negative, decreasing substantially with $T$ reflecting the increased resistivity. The behavior is substantially different for x=0.038 where Fig. 7d, 8, and 9c reveal that $R_S$ is positive between 5 and 20 K, crosses through zero at 25 K, and is strongly negative as $T$ approaches $T_C$=39.5 K. This temperature dependent change in sign of $R_S$ is unusual in that it has not been observed for MnSi for any level of chemical substitution and was verified in crystals grown at a different time having the same Al substitution level [sample s2, Fig 8]. For the $x=0.038$ sample, the temperature dependent change of sign of $R_S$ is fundamentally different from the composition dependent change of sign of $R_S$ that is observed in $Mn_{1-p}Fe_pSi$ and $Mn_{1-r}Co_rSi$ system [34]. Fig 9d presents the variation of $S_H = R_S/\rho_{xx}^2$ as function of $T/T_C$. Consistent with a previous study [32], there is no apparent temperature dependence of $S_H$ for the pure MnSi sample (red). However, a distinct temperature dependence appears as Al or Ga is substituted for Si.

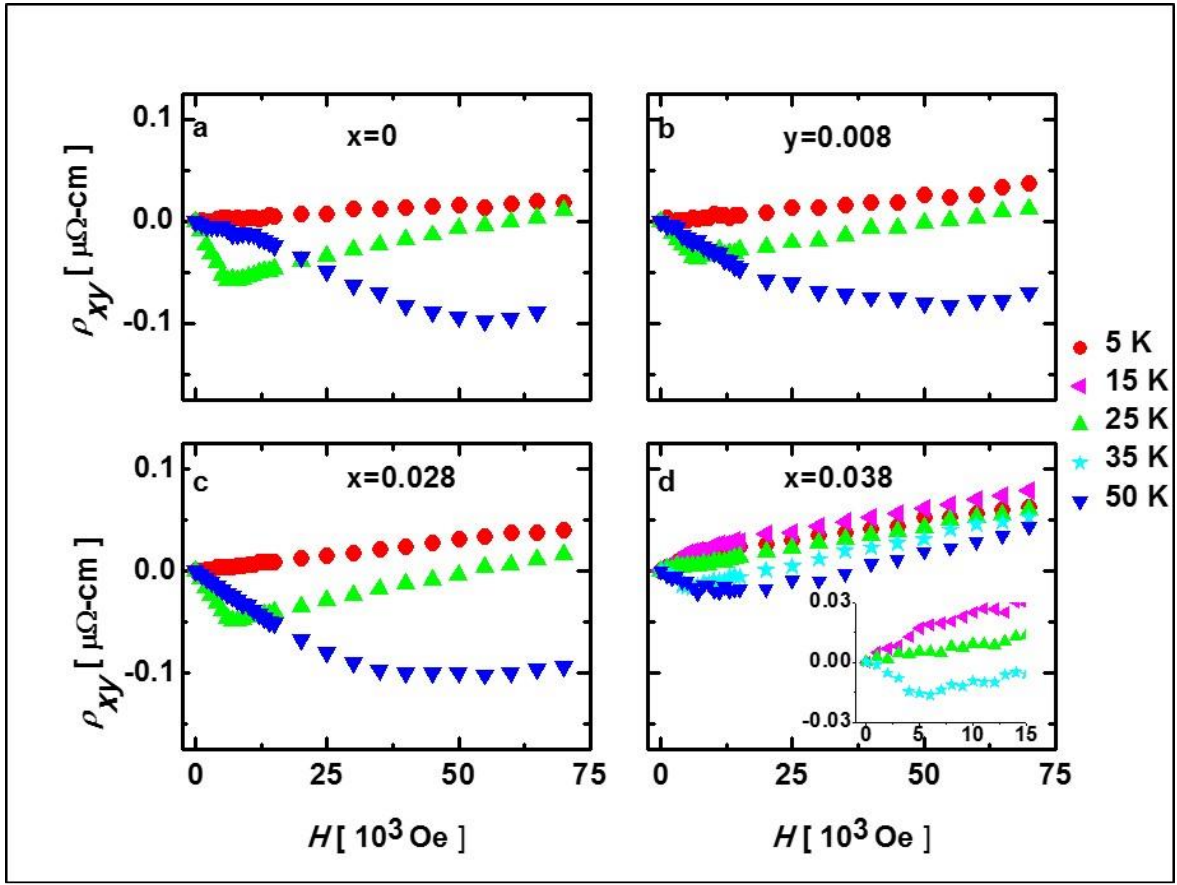

**Fig 7**: Hall resistivity. (a) Hall resistivity, $\rho_{xy}$, for pure MnSi (b) $MnSi_{0.992}Ga_{0.008}$ (c) $MnSi_{0.972}Al_{0.028}$ and (d) $MnSi_{0.962}Al_{0.038}$. Inset: Low field region enlarged to display the change in sign of the anomalous Hall coefficient at low temperature.

### 4. Discussion

This work demonstrates how the magnetic properties of an itinerant magnetic metal, MnSi, can be tuned by simple chemical substitution on a nonmagnetic, Si, site. Naively one should expect very little change in the magnetic moment with chemical substitution on the Si site since the Fermi level lies predominantly within the Mn $d$-bands [36]. However, our results suggest that by about 4% Al substitution for Si the magnetic transition temperature increases by about 11 K and the low temperature saturated moment increases by roughly 20%. These are both significant changes for the relatively small amount of chemical substitutions investigated in this current work. There are two possible causes that we have considered for the changes we observe; (a) a change in carrier density and (b) an increased lattice volume that we characterize as a negative chemical pressure. From our Hall effect measurements, we find a counterintuitive result that the apparent hole carrier density, $n=1/R_0ec$, decreases compared to that of pure MnSi with a significant change at low dopant concentration apparent [Fig 9a]. Given the simple expectation that Al and Ga substitution would reduce the electron density causing an increased hole density, there must be either a more complex change in the band structure from that of the parent compound MnSi, or a change in the relative mobilities of the electron and hole carriers. Studies of the Hall effect in Fe substituted MnSi have also reported a large reduction in the apparent carrier concentration for small substitution levels and noted that the apparent hole concentration in MnSi was larger than expected suggesting a compensating effect of the hole and electron charge carriers [18].

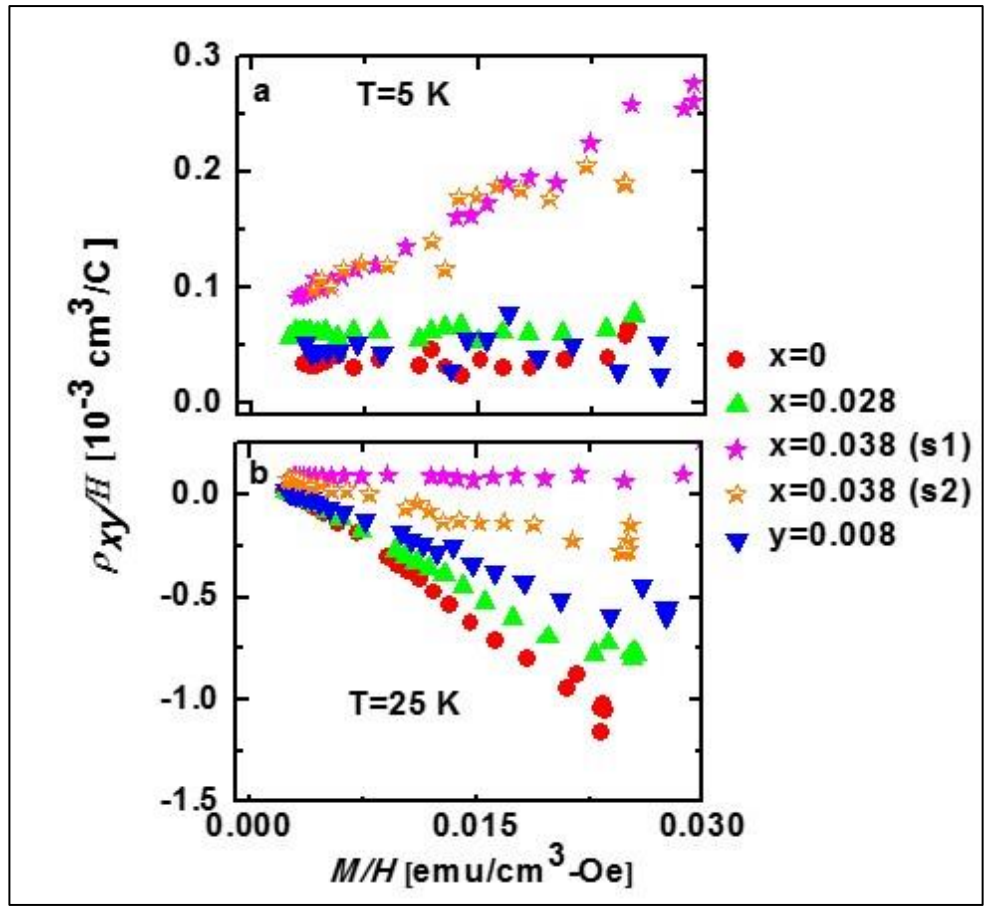

**Fig 8** Standard anomalous Hall effect analysis (a) at 5 K and (b) at 25 K. Hall resistivity, $\rho_{xy}$, divided by magnetic field, $H$, $\rho_{xy}/H$, plotted as a function of the magnetization, $M$, divided by $H$, $M/H$. s1 and s2 represent samples from two different synthesis batches with identical (within error) Al concentration.

For higher Al concentration ($x=0.038$), we observe an unusual temperature dependent sign change in the anomalous contribution for temperatures where the intrinsic contribution is expected to dominate ($T/T_C<1$). The defining feature of the intrinsic anomalous Hall effect is the scattering independence of the anomalous Hall conductivity meaning that $R_S$ can be written as the product of $\rho_{xx}^2$ and a temperature and field independent parameter $S_H$, where $S_H$ is dependent on the Berry curvature found in the electronic structure in proximity to the Fermi surface. This places our observations for MnSi$_{1-x}$Al$_x$ at odds with expectations based upon previous investigations of the Hall effect in *B20* structure materials as demonstrated in Fig 9d [22, 23, 32-34]. To account for a strong temperature dependence to $S_H$, including a change in sign near 25 K, within the accepted description, we would have to postulate a temperature dependent change in the sign of the Berry curvature at temperatures between zero and $T_C = 39.5$ K. This is a very small temperature range over which to expect such a dramatic change to the character of the electronic structure that is usually characterized by energies of order 0.1 eV and calculable only to ~10 meV. Further, we point out that thermal contraction of the crystal lattice upon cooling is an unlikely explanation as only small changes are expected below 100 K and any thermal contraction is likely to reduce the negative chemical pressure brought on by the Al substitution. In contrast, cooling appears to enhance the differences found with respect to nominally pure MnSi.

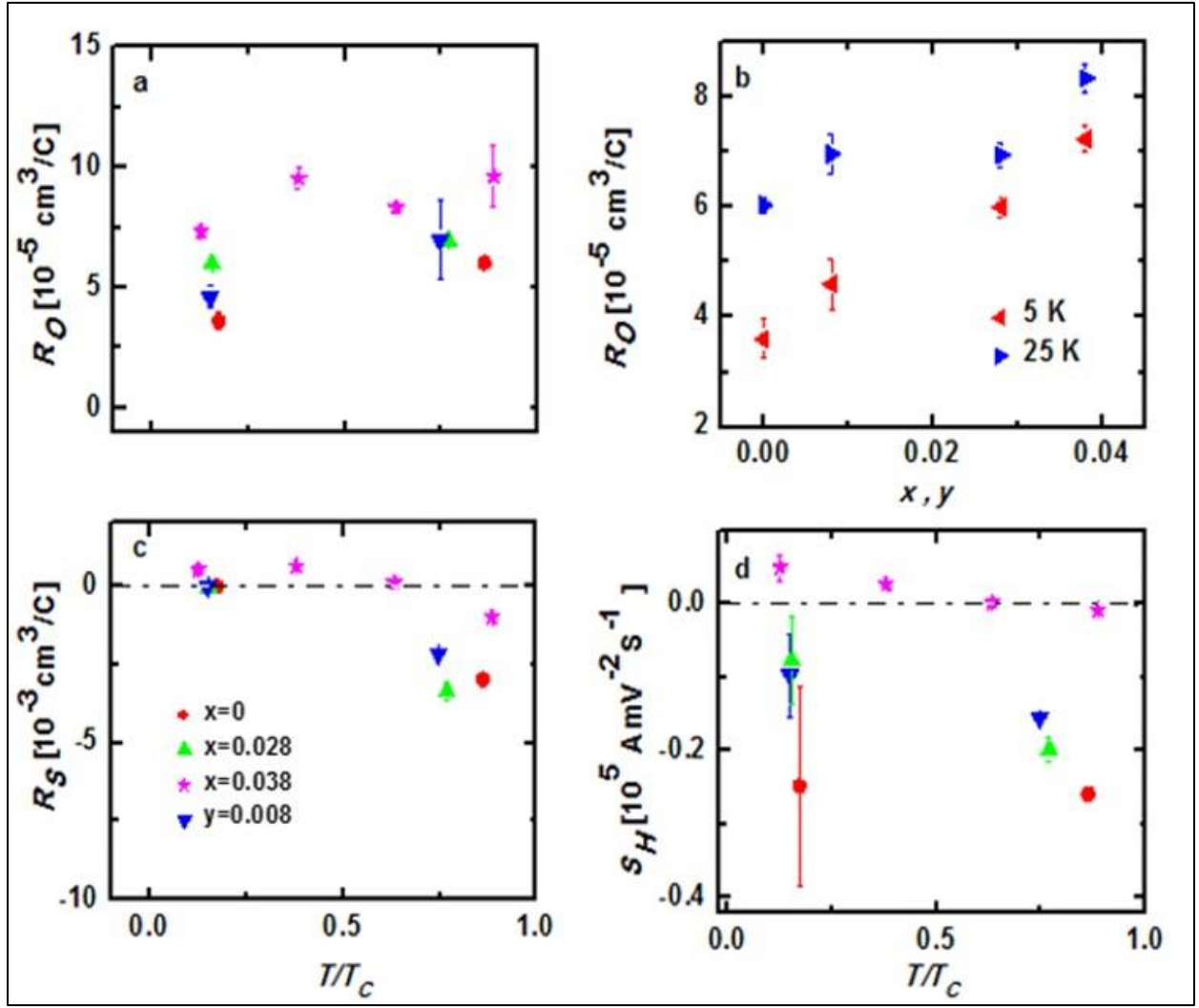

**Fig 9** Hall coefficients: (a) Ordinary Hall coefficient, $R_O$, vs. temperature, $T$, divided by the Curie temperature, $T/T_C$ (b) $R_O$ as a function of dopant concentration at 5 K and 25 K (c) Anomalous Hall coefficient, $R_S$, as a function of $T/T_C$. (d) $S_H$ ($= R_S/\rho_{xx}^2$) as a function of $T/T_C$. Figures (a), (c) and (d) share the same symbols. Fig (a) and (c) share the same x-axis.

In addition, recent theoretical work has suggested that the phase-space Berry curvature can quantitatively account for the DM interaction responsible for the helimagnetism in these materials [37, 38]. Thus the DM interaction determining the helicity of the magnetic state and the AHE may be intimately connected to the details of the band structure which depend upon the spin-orbit coupling strength. Consequently, a change in the sign of $S_H$ with temperature would likely be reflected in the temperature dependence of the helimagnetic period ($\lambda$) or in the chirality (left- or right-handed) of the helimagnetism which can be probed using small angle neutron scattering measurements. The most notable change to MnSi with these low levels of chemical substitutions is the negative chemical pressure created by the expansion of the lattice to accommodate the larger Al and Ga atoms in the structure. We found a very strong relationship between $T_C$ and the unit cell volume. Such a strong relationship is also valid for isoelectronic Ge substitutions, which create only negative chemical pressure without varying the valance electron count [Fig 2a, b]. The changes to $T_C$ and $m_s$ (saturated moment) are likely caused by a significant volume dependence of the exchange correlation resulting from applied pressure as suggested in Ref [8]. Our high-pressure magnetization measurements on two representative samples also indicate that the effect of chemical substitution can be reversed by application of external positive hydrostatic pressure.

We have also explored the effect of negative pressure on the magnetic field dependent magnetic phases of MnSi, including the skyrmion lattice state. The magnetic phase diagram for Ga and Al substituted samples was determined from ac susceptibility measurements and compared with the phase diagram of nominally pure MnSi. The skyrmion lattice phase was observed for all samples, along with the helical, conical, and ferromagnetic field-polarized states observed in MnSi. In fact, the range of temperature and field where the skyrmion phase is stable increases with $x$ and $y$, as compared to MnSi, adding a second method that is distinctly different from a reduction in dimensionality for increasing the stability of this topologically interesting phase [39-41]. A recent study has shown that, in contrast to our measurements, the transition temperature and temperature range for the formation of the *A*-phase decreases due to presence of Mn vacancies and defects [42]. We find no evidence of measureable Mn deficiency in our WDS results. Instead, the increased stability of the skyrmion lattice phase over a wider temperature range is a result of the expanded lattice or from the increased disorder inherent to chemically substituted materials. However, the lack of an increase in $T_C$ or stability of the skyrmion lattice phase with substitutions on the Mn site, suggest that disorder is not the driving factor. The increased field range where this interesting phase is stable suggests a variation in the interaction parameters *D* and *J* [29]. Given the increase in saturation moment and transition temperature, along with the increased unit cell volume, it appears that the spin fluctuations in Al and Ga substituted MnSi may be more localized as compared to nominally pure MnSi, causing an enhanced average moment per Mn site. This suggests that this system progressively varies from weak itinerant to a stronger itinerant behavior with an expansion of the lattice parameter. The increase in $T_C$ and magnetic moment along with increase in the volume of the unit cell can be compared with isostructural MnGe has a much higher $T_C$ (275 K) and a much larger average moment (>1$\mu_B$/Mn) [43].

**Aknowledgements**: The authors acknowledge Dr. Clayton Loehn and Dr. Nele Muttik at Shared Instrumentation Facility (SIF) at LSU for assistance with the chemical analysis. We also thank D. A. Browne and P. T. Sprunger for helpful discussions. This material is based upon work supported by the U.S. Department of Energy under EPSCoR Grant No. DE-SC0012432 with additional support from the Louisiana Board of Regents.

*cdhital@lsu.edu, **ditusa@phys.lsu.edu

**References:**

1. Mühlbauer, S., Binz, B., Jonietz, F., Pfleiderer, C., Rosch, A., Neubauer, A., Georgii, R. and Böni, P., 2009. Skyrmion lattice in a chiral magnet. *Science*, *323*(5916), pp.915-919.
2. Fert, A., Cros, V. and Sampaio, J., 2013. Skyrmions on the track. *Nature nanotechnology*, *8*(3), pp.152-156.
3. Pfleiderer, C., Böni, P., Keller, T., Rößler, U.K. and Rosch, A., 2007. Non-Fermi liquid metal without quantum criticality. *Science*, *316*(5833), pp.1871-1874.
4. Pfleiderer, C., Reznik, D., Pintschovius, L., Löhneysen, H.V., Garst, M. and Rosch, A., 2004. Partial order in the non-Fermi-liquid phase of MnSi. *Nature*, *427*(6971), pp.227-231.
5. Pfleiderer, C., Julian, S.R. and Lonzarich, G.G., 2001. Non-Fermi-liquid nature of the normal state of itinerant-electron ferromagnets. *Nature*, *414*(6862), pp.427-430.
6. Bloch, D., Voiron, J., Jaccarino, V. and Wernick, J.H., 1975. The high field-high pressure magnetic properties of MnSi. *Physics Letters A*, *51*(5), pp.259-261.
7. Yamada, M., Goto, T. and Kanomata, T., 2003. Itinerant metamagnetic properties of MnSi under high pressures. *Physica B: Condensed Matter*, *329*, pp.1136-1137.
8. Panfilov, A.S., 1999. Effect of pressure on magnetic properties of the compound MnSi. *Low Temperature Physics*, *25*, pp.432-435.
9. Yamada, H. and Terao, K., 1999. Itinerant-electron metamagnetism of MnSi at high pressure. *Physical Review B*, *59*(14), p.9342.
10. Miyake, A., Villaume, A., Haga, Y., Knebel, G., Salce, B., Lapertot, G. and Flouquet, J., 2009. Pressure collapse of the magnetic ordering in MnSi via thermal expansion. *Journal of the Physical Society of Japan*, *78*(4), p.044703.
11. Grechnev, G.E., Panfilov, A.S. and Svechkarev, I.V., 1996. Pressure effect on the itinerant magnetism of MnSi and FeSi. *Journal of magnetism and magnetic materials*, *157*, pp.711-712.
12. Moriya, T. and Takahashi, Y., 1984. Itinerant electron magnetism. *Annual Review of Materials Science*, *14*(1), pp.1-25.

13. Thessieu, C., Flouquet, J., Lapertot, G., Stepanov, A.N. and Jaccard, D., 1995. Magnetism and spin fluctuations in a weak itinerant ferromagnet: MnSi. *Solid state communications*, *95*(10), pp.707-712.
14. Ishikawa, Y., Tajima, K., Bloch, D. and Roth, M., 1976. Helical spin structure in manganese silicide MnSi. *Solid State Communications*, *19*(6), pp.525-528.
15. Shanavas, K.V. and Satpathy, S., 2016. Electronic structure and the origin of the Dzyaloshinskii-Moriya interaction in MnSi. *Physical Review B*, *93*(19), p.195101.
16. Collyer, R.D. and Browne, D.A., 2008. Correlations and the Magnetic Moment of MnSi. *Physica B: Condensed Matter*, *403*(5), pp.1420-1422.
17. Bauer, A., Neubauer, A., Franz, C., Münzer, W., Garst, M. and Pfleiderer, C., 2010. Quantum phase transitions in single-crystal Mn $1-$ x Fe x Si and Mn $1-$ x Co x Si: Crystal growth, magnetization, ac susceptibility, and specific heat. *Physical Review B*, *82*(6), p.064404.
18. Manyala, N., Sidis, Y., DiTusa, J.F., Aeppli, G., Young, D.P. and Fisk, Z., 2004. Large anomalous Hall effect in a silicon-based magnetic semiconductor. *Nature materials*, *3*(4), pp.255-262.
19. Manyala, N., DiTusa, J.F., Aeppli, G. and Ramirez, A.P., 2008. Doping a semiconductor to create an unconventional metal. *Nature*, *454*(7207), pp.976-980.
20. Bauer, A. and Pfleiderer, C., 2016. Generic aspects of skyrmion lattices in chiral magnets. In *Topological Structures in Ferroic Materials* (pp. 1-28). Springer International Publishing.
21. Nagaosa, N. and Tokura, Y., 2013. Topological properties and dynamics of magnetic skyrmions. *Nature nanotechnology*, *8*(12), pp.899-911.
22. Do Yi, S., Onoda, S., Nagaosa, N. and Han, J.H., 2009. Skyrmions and anomalous Hall effect in a Dzyaloshinskii-Moriya spiral magnet. *Physical Review B*, *80*(5), p.054416.
23. Everschor-Sitte, K. and Sitte, M., 2014. Real-space Berry phases: Skyrmion soccer. *Journal of Applied Physics*, *115*(17), p.172602.
24. Potapova, N., Dyadkin, V., Moskvin, E., Eckerlebe, H., Menzel, D. and Grigoriev, S., 2012. Magnetic ordering in bulk MnSi crystals with chemically induced negative pressure. *Physical Review B*, *86*(6), p.060406.
25. Sivakumar, K.M., Kuo, Y.K. and Lue, C.S., 2006. Substitutional effect on the transport properties of MnSi. *Journal of Magnetism and Magnetic Materials*, *304*(1), pp.e315-e317.
26. Okada, S., Shishido, T., Ogawa, M., Matsukawa, F., Ishizawa, Y., Nakajima, K., Fukuda, T. and Lundström, T., 2001. MnSi and MnSi $2-$ x single crystals growth by Ga flux method and properties. *Journal of crystal growth*, *229*(1), pp.532-536.
27. See Supplementary materials
28. Nishihara, Y., Waki, S. and Ogawa, S., 1984. Mössbauer study of Mn $1-$ x Fe x Si in external magnetic fields. *Physical Review B*, *30*(1), p.32.
29. Grigoriev, S.V., Maleyev, S.V., Okorokov, A.I., Chetverikov, Y.O., Böni, P., Georgii, R., Lamago, D., Eckerlebe, H. and Pranzas, K., 2006. Magnetic structure of MnSi under an applied field probed by polarized small-angle neutron scattering. *Physical Review B*, *74*(21), p.214414.

30. Nagaosa, N., Sinova, J., Onoda, S., MacDonald, A.H. and Ong, N.P., 2010. Anomalous hall effect. *Reviews of modern physics*, *82*(2), p.1539.
31. Neubauer, A., Pfleiderer, C., Binz, B., Rosch, A., Ritz, R., Niklowitz, P.G. and Böni, P., 2009. Topological Hall effect in the A phase of MnSi. *Physical review letters*, *102*(18), p.186602.
32. Lee, M., Onose, Y., Tokura, Y. and Ong, N.P., 2007. Hidden constant in the anomalous Hall effect of high-purity magnet MnSi. *Physical Review B*, *75*(17), p.172403.
33. Glushkov, V.V.E., Lobanova, I.I., Ivanov, V.Y. and Demishev, S.V.E., 2015. Anomalous Hall effect in MnSi: Intrinsic to extrinsic crossover. *JETP Letters*, *101*(7), pp.459-464.
34. Franz, C., Freimuth, F., Bauer, A., Ritz, R., Schnarr, C., Duvinage, C., Adams, T., Blügel, S., Rosch, A., Mokrousov, Y. and Pfleiderer, C., 2014. Real-space and reciprocal-space berry phases in the Hall effect of Mn $1-x$ Fe $x$ Si. *Physical review letters*, *112*(18), p.186601.
35. Chapman, B.J., Grossnickle, M.G., Wolf, T. and Lee, M., 2013. Large enhancement of emergent magnetic fields in MnSi with impurities and pressure. *Physical Review B*, *88*(21), p.214406.
36. Jeong, T. and Pickett, W.E., 2004. Implications of the B20 crystal structure for the magnetoelectronic structure of MnSi. *Physical Review B*, *70*(7), p.075114.
37. Freimuth, F., Bamler, R., Mokrousov, Y. and Rosch, A., 2013. Phase-space Berry phases in chiral magnets: Dzyaloshinskii-Moriya interaction and the charge of skyrmions. *Physical Review B*, *88*(21), p.214409.
38. Gayles, J., Freimuth, F., Schena, T., Lani, G., Mavropoulos, P., Duine, R.A., Blügel, S., Sinova, J. and Mokrousov, Y., 2015. Dzyaloshinskii-Moriya Interaction and Hall Effects in the Skyrmion Phase of Mn $1-x$ Fe $x$ Ge. *Physical review letters*, *115*(3), p.036602.

39. Zhang, S.L., Chalasani, R., Baker, A.A., Steinke, N.J., Figueroa, A.I., Kohn, A., van der Laan, G. and Hesjedal, T., 2016. Engineering helimagnetism in MnSi thin films. *AIP Advances*, *6*(1), p.015217.
40. Liang, D., DeGrave, J.P., Stolt, M.J., Tokura, Y. and Jin, S., 2015. Current-driven dynamics of skyrmions stabilized in MnSi nanowires revealed by topological Hall effect. *Nature communications*, *6*.
41. Yu, X.Z., Kanazawa, N., Onose, Y., Kimoto, K., Zhang, W.Z., Ishiwata, S., Matsui, Y. and Tokura, Y., 2011. Near room-temperature formation of a skyrmion crystal in thin-films of the helimagnet FeGe. *Nature materials*, *10*(2), pp.106-109.
42. Ou-Yang, T.Y., Shu, G.J., Lin, J.Y., Hu, C.D. and Chou, F.C., 2015. Mn vacancy defects, grain boundaries, and A-phase stability of helimagnet MnSi. *Journal of Physics: Condensed Matter*, *28*(2), p.026004.
43. DiTusa, J.F., Zhang, S.B., Yamaura, K., Xiong, Y., Prestigiacomo, J.C., Fulfer, B.W., Adams, P.W., Brickson, M.I., Browne, D.A., Capan, C. and Fisk, Z., 2014. Magnetic, thermodynamic, and electrical transport properties of the noncentrosymmetric B 20 germanides MnGe and CoGe. *Physical Review B*, *90*(14), p.144404.

# Supplementary Materials

Effect of Negative Pressure on the Prototypical Itinerant Magnet MnSi

C. Dhital, M. A. Khan, M. Saghayezhian, W. A. Phelan, D. P. Young, R. Y. Jin, and J. F. DiTusa

Department of Physics and Astronomy, Louisiana State University, Baton Rouge, LA 70803

# 1. Determination of Transition Temperature $T_C$

The Curie temperature, $T_C$, was determined by taking the first derivative of magnetic susceptibility $\chi$ [Fig. 1a in main text] with respect to temperature, $T$ as shown in Fig S1. $T_C$ is defined as the midpoint of the decreasing $d\chi/dT$ curve as shown in the figure.

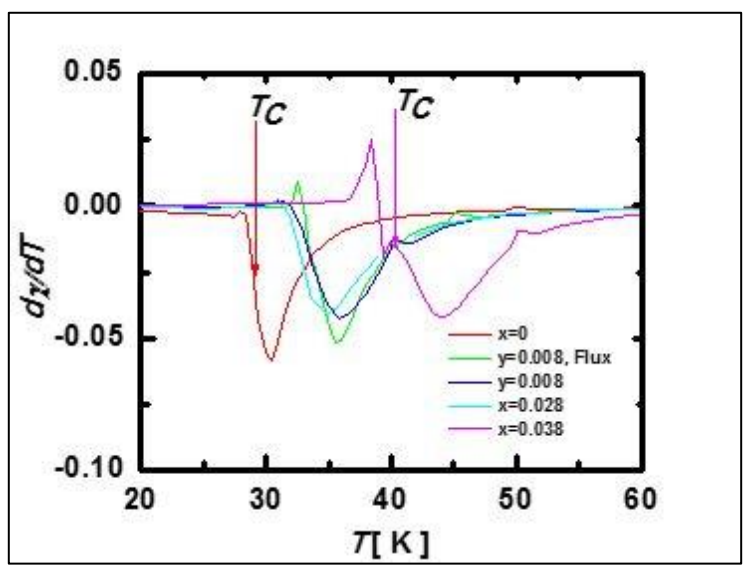

**Fig S1** Determination of Curie temperatures: Variation of derivative of magnetic susceptibility $\chi$, $d\chi/dT$ with respect to temperature, $T$.

# 2. X-ray Photoemission study

Monochromatized X-ray Photoemission Spectroscopy (XPS) was performed to observe changes in the Mn oxidation state due to chemical substitution. The samples were sputtered with $Ar^+$ ions for 1 hr before measurements. The data were collected at 300 K at normal emission angle thereby probing the bulk of the sample. Only minor changes to the spectra are observed between 640 and 645 eV implying no significant change in oxidation state or the correlation strength.

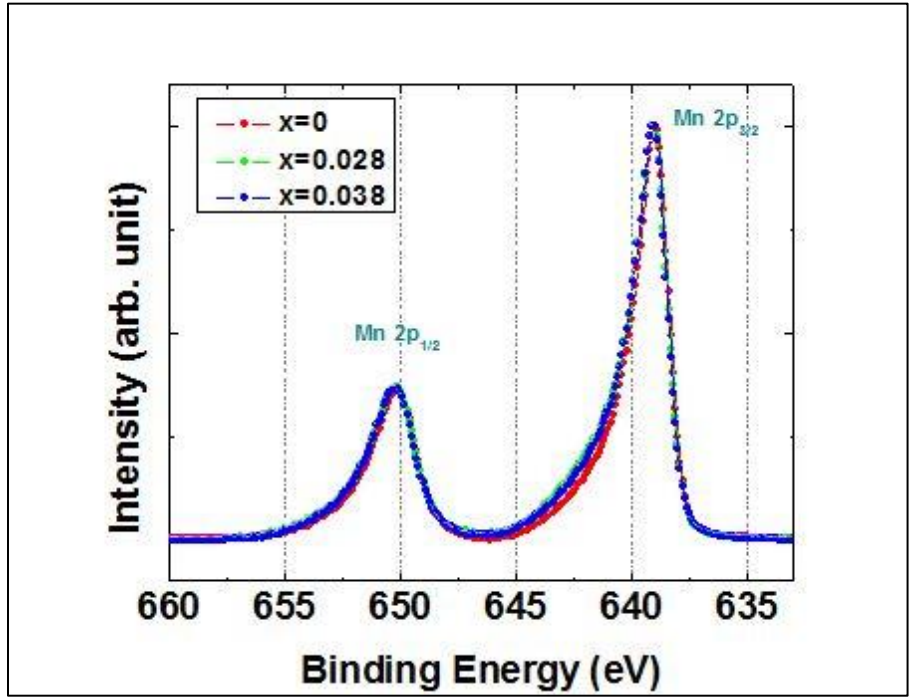

**Fig S2**: XPS spectra showing Mn 2P edge for pure and Al substituted samples, $MnSi_{1-x}Al_x$.

# 3. Determination of fields $H_1, H_2, H_3, H_4$ from ac susceptibility measurements.

The value of the critical fields shown in the phase diagram (Fig 5 in main text) were obtained from following ac susceptibility measurements.

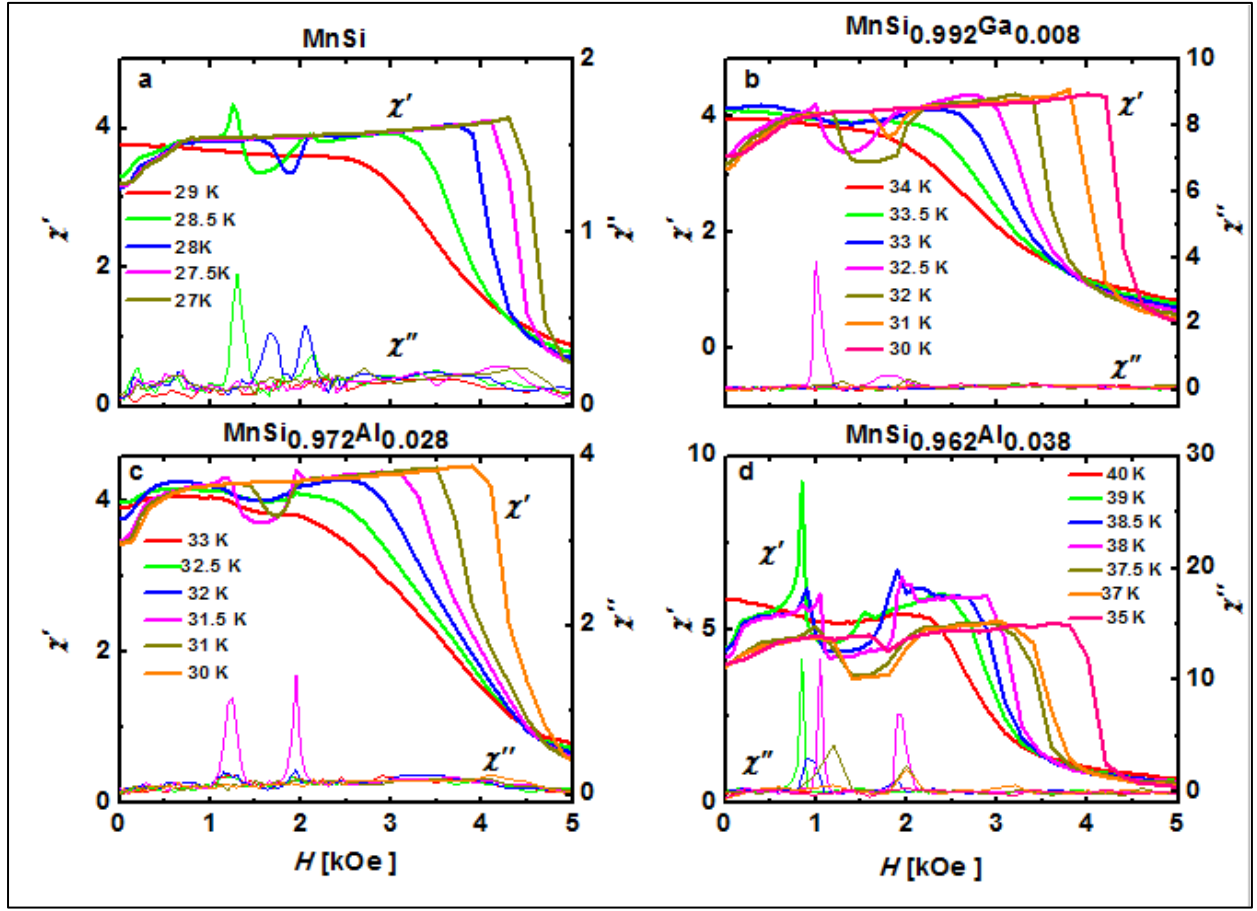

**Fig S3**: ac susceptibility as function of dc field for (a) MnSi (b) MnSi$_{0.992}$Ga$_{0.008}$ (c) MnSi$_{0.972}$Al$_{0.028}$ and (d) MnSi$_{0.962}$Al$_{0.038}$. The measurements were taken at an ac amplitude of 1 Oe at 100 Hz frequency. In each figure, the upper set of curves represents the real part ($\chi'$) (left axis) and the lower set represents the imaginary part ($\chi''$) (right axis).

## 4. Tabulated lattice constant and magnetic data

The list of Ga and Al doped samples along with their effective Curie moment, Weiss temperature, saturated moment, and transition temperature are presented in Table ST1. Al and Ga substituted samples are single crystals (SC) with chemical substitution levels determined via chemical analysis (see experimental details section of the main paper). Polycrystalline (PC) Ge and Fe substituted samples, prepared to compare with our Curie temperature and lattice constant results, are also included.

**Table ST1**

| Sample | $a$ (Å) | $T_C$ (K) | $m_{eff}$ (μB/FU) | Saturated moment ($M$) at 5 K (μB/FU) |
|---|---|---|---|---|
| **MnSi (SC)** | 4.5613 (3) | 29±0.25 | 2.22 ±0.02 | 0.42±0.01 |
| **MnSi$_{0.992}$Ga$_{0.008}$(SC)** | 4.5639 (4) | 33±0.25 | 2.25 ±0.02 | 0.45±0.01 |
| **MnSi$_{0.972}$Al$_{0.028}$ (SC)** | 4.5645 (2) | 32.5±0.25 | 2.36±0.02 | 0.44±0.01 |
| **MnSi$_{0.962}$Al$_{0.038}$ (SC)** | 4.5688 (6) | 39.5±0.25 | 2.35±0.02 | 0.52±0.01 |
| **MnSi$_{0.98}$Ge$_{0.02}$ (PC)** | 4.5626 (2) | 30.5 ±0.25 | 2.35±0.02 | 0.43±0.01 |
| **MnSi$_{0.96}$Ge$_{0.04}$ (PC)** | 4.5631 (2) | 31.5±0.25 | 2.34±0.02 | 0.43±0.01 |
| **Mn$_{0.95}$Fe$_{0.05}$Si (PC)** | 4.5549 (4) | 13.5±0.25 | 2.11±0.02 | 0.33±0.01 |
| **Mn$_{0.9}$Fe$_{0.1}$Si (PC)** | 4.5518 (3) | 6.5±0.25 | 1.93±0.02 | 0.23±0.01 |